\documentclass[12pt]{article}
\usepackage{graphicx}

\newcommand{\initiate}{\setcounter{equation}{0}}
\usepackage{amsmath}
\usepackage{amsbsy}
\usepackage{amsfonts}
\usepackage{amsthm}
\begin{document}

\title{Q-stars in anti de Sitter spacetime}

\author{Athanasios Prikas}

\date{}

\maketitle

Physics Department, National Technical University, Zografou
Campus, 157 80 Athens, Greece.\footnote{e-mail:
aprikas@central.ntua.gr}

\begin{abstract}
We study q-stars with various symmetries in anti de Sitter spacetime in
$3+1$ dimensions. Comparing with the case of flat spacetime, we
find that the value of the field at the center of the soliton is
larger when the other parameters show a more complicated
behavior. We also investigate their phase space when the symmetry
is local and the effect of the charge to its stability.
\end{abstract}

\vspace{1em}

PACS number(s): 11.27.+d, 04.40.-b

\newpage

\section{Introduction}

Recently, a great amount of theoretical interest has been focused
on anti de Sitter (AdS) spacetime, due to the close relation
between (gravitating) fields within AdS spacetime and a field
theory at the boundary of the above spacetime
\cite{witten,maldacena}. String-inspired theories work mainly in
the framework of a spacetime with negative curvature \cite{string
review}, rising the need to study fields within this framework.

Scalar fields coupled to gravity investigated in the work of Kaup,
\cite{kaup}, and Ruffini and Bonazzola \cite{ruf}, considering a
scalar field with no self-interactions, and in other works,
considering a scalar field with self interactions,
\cite{quartic1,quartic2}, or scalar fields with local symmetries
\cite{charged}. Other gravity theories have been also taken into
account in a series of papers \cite{tor1,tor2,tor3,tor4}. The
gravitating objects arising form such actions, the stars, are
sometimes called ``mini boson stars" due to their small relative
magnitude.

Soliton stars are stable field configurations which deserve their
stability even in the absence of gravity and remain as
non-topological solitons. One subclass of soliton stars,
\cite{fr1,fr2,fr3,brito}, were based on the earlier work of
Friedberg, Lee and Sirlin, are known as ``large boson stars".
Another class of non-topological soliton stars, q-stars, appeared
as relativistic generalizations of q-balls. Q-balls are
non-topological solitons in Lagrangians of highly non-linearly
self-interacting boson fields with a global $U(1)$, $SU(3)$ or
$SO(3)$ symmetry \cite{col1,col2,tracas,kusenco}, or a local
$U(1)$, \cite{local}. The scalar field rotates in its internal
$U(1)$ space with a frequency equal to
$\sqrt{U/|\phi|^2}_{\textrm{min}}$ and takes the special value
that minimizes the above quantity. Q-balls concentrated special
interest, due to their role in the flat-directions baryogenesis in
supersymmetric extensions of Standard Model. Q-stars are
relativistic extensions of q-balls, with one or two scalar fields
and a global, \cite{qstars-global}, or local,
\cite{qstars-charged}, $U(1)$ symmetry, non-abelian symmetry,
\cite{qstars-nonabelian}, or with fermions and a scalar field,
\cite{qstars-fermion}.

There are some relations governing the three types of boson stars
and revealing their crucial differences. The following
estimations hold for mass and radius of the ``mini", q- and
``large" boson stars respectively (in $3+1$ dimensions):
\begin{equation}\label{1.1}
M_{\textrm{mini}}\sim \frac{M^2_{\textrm{Pl}}}{m}, \hspace{1em}
M_{\textrm{q}}\sim\frac{M_{\textrm{Pl}}^3}{m^2}, \hspace{1em}
M_{\textrm{large}}\sim\frac{M_{\textrm{Pl}}^4}{m^3}\ ,
\end{equation}
\begin{equation}\label{1.2}
R_{\textrm{mini}}\sim m^{-1}, \hspace{1em}
R_{\textrm{q}}\sim\frac{M_{\textrm{Pl}}}{m^2}, \hspace{1em}
R_{\textrm{large}}\sim\frac{M_{\textrm{Pl}}^2}{m^3}\ ,
\end{equation}
where $m$ is a mass scale, usually the mass of the free scalar
field. The above estimations justify the names ``mini" and
``large".

In the present article we investigate the formation of q-stars
with one and two scalar fields, q-stars with non-abelian symmetry
and scalar-fermion q-stars. We solve numerically the coupled
Einstein-Lagrange equations and investigate their properties,
radius, field value at the origin, total mass and particle
number. We verify their stability with respect to fission into
free particles and discuss especially the influence of the
cosmological constant to the soliton properties. When the
spacetime is flat, the results of
\cite{qstars-global,qstars-charged,qstars-nonabelian,qstars-fermion}
are reproduced. We compare our results with the behavior of boson
(not soliton) stars in the negative curvature spacetime
background \cite{astefanesei}. We finally describe charged scalar
fields coupled to gravity and check the stability of the solitons
with respect to free particles decay, because when the symmetry
is local the total energy of the star is larger.

\initiate
\section{Q-stars with one scalar field}

We consider a static, spherically symmetric metric:
\begin{equation}\label{2.1}
ds^2=-e^{\nu}dt^2+e^{\lambda}d{\rho}^2+{\rho}^2d{\alpha}^2+{\rho}^2\sin^2\alpha
d{\beta}^2\ ,
\end{equation}
with $g_{tt}=-e^{\nu}$. The scalar field is supposed to form a
spherically symmetric configuration. A time dependence ansatz,
suitable for minimizing the energy, is:
\begin{equation}\label{2.2}
\phi(\vec{\rho},t)=\sigma(\rho)e^{-\imath\omega t}\ .
\end{equation}
The action in natural units for a scalar field coupled to gravity
in $3+1$ dimensions is:
\begin{equation}\label{2.3}
S=\int_{\mathcal{M}} d^4x\sqrt{-g}\left[\frac{R-2\Lambda}{16\pi
G}+g^{\mu\nu}
{({\partial}_{\mu}\phi)}^{\ast}({\partial}_{\nu}\phi)-U\right]+\frac{1}{8\pi
G}\int_{\partial\mathcal{M}}d^3x\sqrt{-h}K \ ,
\end{equation}
where the second term is the Hawking-Gibbons surface term
\cite{gibbons}, $\Lambda$ stands for the cosmological constant
regarded here to be negative or zero as a limiting case,
$\partial\mathcal{M}$ is the boundary of the space $\mathcal{M}$,
$R$ and $K$ are the curvature scalars within the space and at its
boundary respectively and $g$ and $h$ stand for the metric
determinant within the space and at the boundary respectively.
The second term does not affect the classical equations of motion
of the matter fields but appears when the quantization is under
investigation. The energy-momentum tensor is:
\begin{equation}\label{2.4}
T_{\mu\nu}={({\partial}_{\mu}\phi)}^{\ast}({\partial}_{\nu}\phi)+
({\partial}_{\mu}\phi){({\partial}_{\nu}\phi)}^{\ast}
-g_{\mu\nu}[g^{\alpha\beta}{({\partial}_{\alpha}\phi)}^{\ast}({\partial}_{\beta}\phi)]
-g_{\mu\nu}U\ .
\end{equation}
We will later choose a simple potential, admitting q-ball type
solutions in the absence of gravity. This potential, and,
consequently, the whole theory need not be renormalizable but an
effective one. The Euler-Lagrange equation for the matter field
is:
\begin{equation}\label{2.5}
\left[1/\sqrt{|g|}{\partial}_{\mu}(\sqrt{|g|}g^{\mu\nu}{\partial}_{\nu})-
\frac{dU}{d{|\phi|}^2}\right]\phi=0\ ,
\end{equation}
taking now the form:
\begin{equation}\label{2.6}
{\sigma}''+[2/\rho+(1/2)({\nu}'-{\lambda}')]{\sigma}'+e^{\lambda}
{\omega}^2e^{-\nu} \sigma
-e^{\lambda}\frac{dU}{d{\sigma}^2}\sigma=0\ .
\end{equation}
The Einstein equations are:
\begin{equation}\label{2.7}
G_{\ \nu}^{\mu}\equiv R_{\ \nu}^{\mu}-\frac{1}{2}{\delta}_{\
\nu}^{\mu}R=8\pi G T_{\ \nu}^{\mu}-\Lambda{\delta}_{\ \nu}^{\mu}
\ ,
\end{equation}
and with the assumptions of eqs. \ref{2.1}, \ref{2.2} the two
independent of them, $G^0_{\ 0}$ and $G^1_{\ 1}$ take the
following form, respectively:
\begin{equation}\label{2.8}
\frac{e^{-\lambda}-1}{{\rho}^2}-e^{-\lambda}\frac{\lambda'}{\rho}=8\pi
G(-W-V-U)-\Lambda\ ,
\end{equation}
\begin{equation}\label{2.9}
\frac{e^{-\lambda}-1}{{\rho}^2}+e^{-\lambda}\frac{\nu'}{\rho}=8\pi
G(W+V-U)-\Lambda\ ,
\end{equation}
where
\begin{equation}\label{2.10}
\begin{split}
W&\equiv e^{-\nu}{\left(\frac{\partial\phi}{\partial
t}\right)}^{\ast}\left(\frac{\partial\phi}{\partial t}\right)=
e^{-\nu}{\omega}^2{\sigma}^2\ , \\ V&\equiv
e^{-\lambda}{\left(\frac{\partial\phi}{\partial\rho}\right)}^{\ast}
\left(\frac{\partial\phi}{\partial\rho}\right)=
e^{-\lambda}{\sigma'}^2\ .
\end{split}
\end{equation}

There is a Noether current due to the global $U(1)$ symmetry of
the Lagrangian:
\begin{equation}\label{2.11}
j^{\mu}=\sqrt{-g}g^{\mu\nu}\imath({\phi}^{\ast}{\partial}_{\nu}\phi
-\phi{\partial}_{\nu}{\phi}^{\ast})\ .
\end{equation}
The current is conserved according to the equation:
\begin{equation}\label{2.12}
j^{\mu}_{\ ;\mu}=0\ .
\end{equation}
The total charge is defined as:
\begin{equation}\label{2.13}
Q=\int d^3xj^0\ ,
\end{equation}
now taking the form
\begin{equation}\label{2.14}
Q=8\pi\int\rho^2 d\rho\omega{\sigma}^2e^{-\nu/2}e^{\lambda/2}\ .
\end{equation}

If each particle is assigned with a unity charge then the total
charge equals to the particle number of the configuration.
Rescaling the Lagrangian parameters in a specific way we will
describe later, we find that in our new units the mass is unity
and, consequently, the energy of the free particles with the same
charge equals to the total charge. So, the total charge
determines the stability of the soliton with respect to decay
into free particles.

We define:
\begin{equation}\label{2.15}
A\equiv e^{-\nu}\ , \hspace{1em} B\equiv e^{-\lambda}
\end{equation}
and rescale:
\begin{equation}\label{2.16}
\begin{split}
\tilde{\rho}=\rho m\ , \hspace{1em} \tilde{\omega}&=\omega/m\ ,
\hspace{1em} \tilde{\phi}=\phi/m \, \\ \widetilde{U}=U/m^4\ ,
\hspace{1em} \widetilde{W}&=W/m^4\ , \hspace{1em}
\widetilde{V}=V/m^4\ ,
\end{split}
\end{equation}
where $m$ is the mass of the free scalar field and can be a
general mass scale. We also make the redefinitions:
\begin{equation}\label{2.17}
\widetilde{\Lambda}\equiv \frac{\Lambda}{8\pi Gm^4}\ ,
\hspace{1em} \tilde{r}\equiv \epsilon\tilde{\rho}\ ,
\end{equation}
where $\epsilon$ is a very useful quantity in the study of
gravitating bosonic field configurations, defined as:
\begin{equation}\label{2.18}
\epsilon\equiv \sqrt{8\pi Gm^2}\ .
\end{equation}
When $m$ is of some $GeV$, then the above quantity is extremely
small, and quantities of the same order of magnitude can be
neglected. The second part of eq. \ref{2.17} summarizes the
effect of gravity upon to the soliton, because gravity, according
to Schwarzschild condition, becomes important when:
$$R\sim GM,$$ where $R$ is the soliton radius and $M$ its total mass.
The energy density within the soliton is approximately constant
and equal to $m^4$ (or ${\phi}^4$). So, one can find that, in
$3+1$ dimensions:
\begin{equation}\label{2.19}
R\sim{\epsilon}^{-1}\ ,
\end{equation}
and, consequently, $\tilde{r}\sim1$, simplifying considerably the
numerical procedures. Inside the soliton both the metric and
matter fields variate extremely slowly with respect to the
radius. This means that the matter field starts with an initial
value ${\phi}_0$ at the center of the soliton and ends with the
zero value at the tail, \textit{without oscillations meanwhile}.
This intuitive argument reflects to the relation:
$V\propto{\epsilon}^2$, resulting from eqs.
\ref{2.16}-\ref{2.18}. Einstein equations take the form, ignoring
the $O(\epsilon)$ quantities and dropping the tildes:
\begin{equation}\label{2.20}
\frac{1-A}{r^2}-\frac{1}{r}\frac{dA}{dr}={\omega}^2{\sigma}^2B+U+\Lambda\
,
\end{equation}
\begin{equation}\label{2.21}
\frac{A-1
}{r^2}-\frac{1}{r}\frac{A}{B}\frac{dB}{dr}={\omega}^2{\sigma}^2B-U-\Lambda\
,
\end{equation}
and the Euler-Lagrange equation is:
\begin{equation}\label{2.22}
{\omega}^2B-\frac{dU}{d{\sigma}^2}=0\ .
\end{equation}

A rescaled potential admitting q-ball type solutions in the
absence of gravity is:
\begin{equation}\label{2.23}
U=a{|\phi|}^2-b{|\phi|}^3+c{|\phi|}^4\ ,
\end{equation}
and eq. \ref{2.22} gives:
\begin{equation}\label{2.24}
|\phi|=\left(\frac{3}{2}b+\sqrt{\frac{9}{4}b^2-4bc(a-{\omega}^2B)}\right)/(4c)\
,
\end{equation}
where $a$, $b$ and $c$ are all positive quantities. Another
potential we will use here is:
\begin{equation}\label{2.25}
U=a{|\phi|}^2-b{|\phi|}^4+c{|\phi|}^6=a{\sigma}^2-b{\sigma}^4+c{\sigma}^6
\end{equation}
for which eq. \ref{2.22} gives:
\begin{equation}\label{2.26}
{\phi}^2=\left(b+\sqrt{b^2-3c(a-{\omega}^2B)}\right)/(3c)\ .
\end{equation}
If we choose for simplicity $a=b=1$, $c=1/3$ the potential takes
the simple form:
\begin{equation}\label{2.27}
U=\frac{1}{3}(1+{\omega}^3B^{3/2})\ .
\end{equation}

\begin{figure}
\centering
\includegraphics{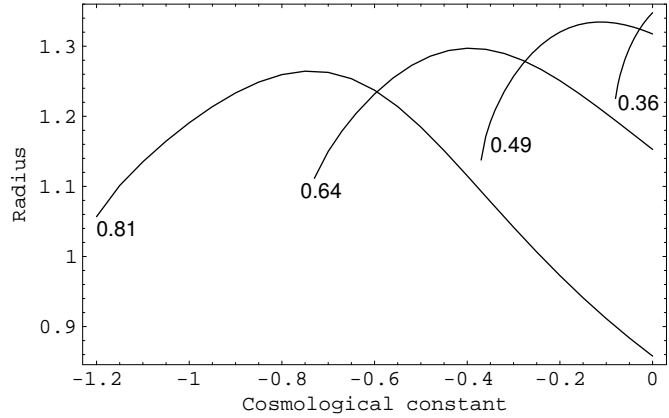}
\caption{The radius for a q-star with one scalar field as a
function of the cosmological constant for four different values
of $A_{\textrm{sur}}$, or, equivalently, $\omega$ obtained from
eq. \ref{2.30}. Small $A_{\textrm{sur}}$ indicates strong gravity
at the surface and in general at the region where the soliton is.}
\label{figure1.1}
\end{figure}

\begin{figure}
\centering
\includegraphics{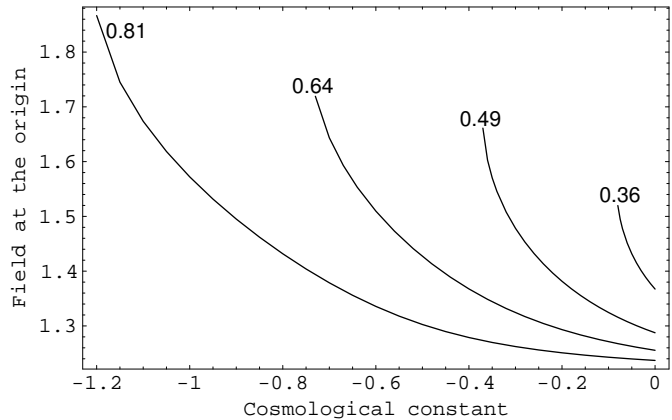}
\caption{The value of the scalar field at the center of the
soliton for a q-star with one scalar field as a function of the
cosmological constant for four values of $A_{\textrm{sur}}$.}
\label{figure1.2}
\end{figure}

\begin{figure}
\centering
\includegraphics{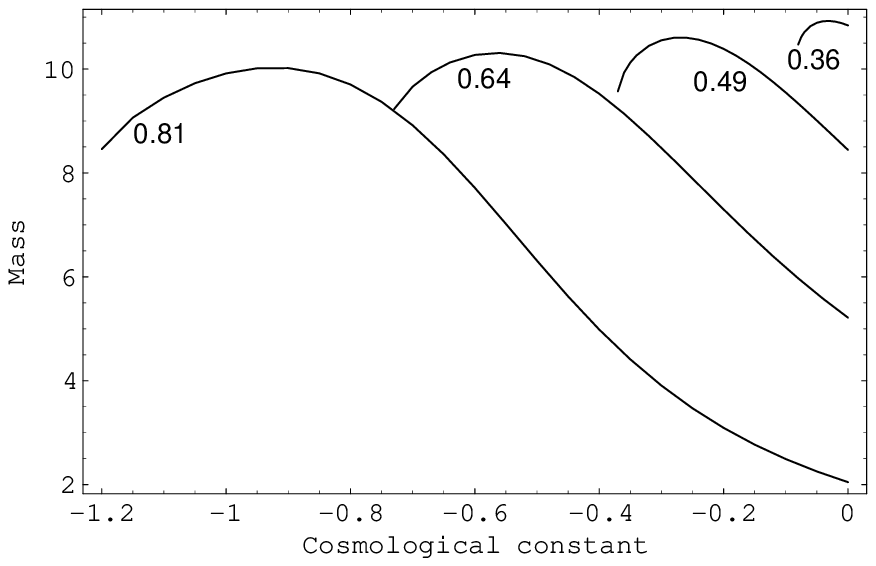}
\caption{The total mass for a q-star with one scalar field as a
function of the cosmological constant for four different values
of $A_{\textrm{sur}}$.} \label{figure1.3}
\end{figure}

\begin{figure}
\centering
\includegraphics{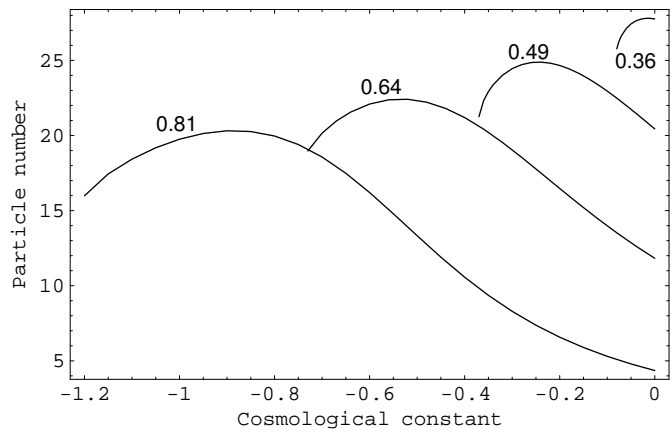}
\caption{The particle number for a q-star with one scalar field as
a function of the cosmological constant for four different values
of $A_{\textrm{sur}}$.} \label{figure1.4}
\end{figure}

The surface width is of order of $m^{-1}$. Within this, the matter
field varies rapidly, from a value ${\sigma}$ at the inner edge of
the surface, to zero at the outer, but the metric fields not. So,
dropping from the Lagrange equation the $O(\epsilon)$ quantities
we take:
\begin{equation}\label{2.28}
\frac{\delta(W-U-V)}{\delta\sigma}=0\ .
\end{equation}
The above equation can be straightforward integrated and, because
all energy quantities are zero at the outer edge of the surface,
the result gives the following equation, holding true only within
the surface:
\begin{equation}\label{2.29}
V+U-W=0\ .
\end{equation}
At the inner edge of the surface $\sigma'$ is zero in order to
match the interior with the surface solution. So, at the inner
edge of the surface the equality $W=U$ gives together with eqs.
\ref{2.26}, \ref{2.27}:
\begin{equation}\label{2.30}
\omega=\frac{A_{\textrm{sur}}^{1/2}}{2}=\frac{B_{\textrm{sur}}^{-1/2}}{2}\
.
\end{equation}
Eq. \ref{2.30} is the eigenvalue equation for the frequency of
the q-star, relating straightforward a feature of its internal
$U(1)$ space with the spacetime curvature, as this is measured by
$A_{\textrm{sur}}$. So, either $A_{\textrm{sur}}$, or $\omega$
can be alternatively used as the measure of the total gravity
strength.

Because ${\sigma}_{\textrm{in}}$ is the mean value of the scalar
field in the soliton interior, its mean value within the surface
is approximately ${\sigma}_{\textrm{in}}/2$. So, the energy
density, equal to the sum of $W$, $U$, and $V$, has the same
order of magnitude either in the interior, or within the surface.
The ratio of the energy amount stored in the interior to that
stored in the surface is $\sim R^3m^4/(m^{-1}R^2m^4)\sim
m^{-1}{\epsilon}^{-1}$, so the energy stored in the surface is
negligible, when compared to the energy stored in the interior.
The same discussion holds true for the charge contribution. The
mass of the star can be calculated by using the $T^0_{\ 0}$
component of the energy-momentum tensor which gives:
\begin{equation}\label{2.31}
E=4\pi\int_0^Rdrr^2\left({\omega}^2{\sigma}^2B+\frac{1}{3}(1+{\omega}^3B^{3/2})\right)\
,
\end{equation}
where, according to the above discussion, we integrate within the
soliton interior, ignoring the surface contribution. Equivalently
we may use the the proper Schwarzschild formula:
$$A=1-\frac{2GM(\rho)}{\rho^{D-3}}-\frac{2\Lambda\rho^2}{(D-2)(D-1)}\
,$$ where $D$ is the dimensionality of the spacetime, here $4$,
and $M(r)$ represents the total mass of the field configuration
(in $3+1$ dimensions), when $\rho\rightarrow\infty$, as a self
consistency check of our calculations. (In order to apply
correctly the above relation, we should keep in mind the
rescalings of eqs. \ref{2.16}-\ref{2.17}, and the fact that if
$\mathcal{E}$ is the energy density, then
$\widetilde{\mathcal{E}}=\mathcal{E}/m^4$ and $M=\int_0^\infty
d^3\rho\mathcal{E}$. Then, we take from the above relation and
for the rescaled quantities: $M= 4\pi r(1-A(r)-(1/3)\Lambda r^2)$.

At the exterior the Einstein equations can be solved analytically:
\begin{equation}\label{2.32}
A(r)=\frac{3RA_{\textrm{sur}}+3(r-R)-\Lambda(r^3-R^3)}{3r}\ ,
\hspace{1em} B(r)=1/A(r)\ ,
\end{equation}
with eq. \ref{2.32} being a formula alternative to the
Schwarzschild one presented above.

We use a fourth order Runge-Kutta scheme to solve Einstein
equations. The star radius is in $(\epsilon m)^{-1}$ units, the
total mass in $(\epsilon^{3/2}m^{-7/2})^{-1}$ units, and the
charge is in $(\epsilon^{3/2}m^{-5/2})^{-1}$ units. From figures
\ref{figure1.1}-\ref{figure1.4} we see that for small values of
the cosmological constant the soliton radius, mass and particle
number increase with the increase of the cosmological constant in
absolute values. From an intuitive point of view, negative
cosmological constant reflects a competitive effect to gravity
attraction. So, a large soliton increases its total mass and
consequently its particle number and radius, in order to be
stable against this ``negative" gravity implied by the negative
cosmological constant. But, when $\Lambda$ exceeds a certain
value, no additional energy amount can deserve the soliton
stability, \textit{if it is too extended}. So, when $\Lambda$
exceeds this certain value, the star shrinks, and, consequently,
its energy and charge decreases. Also, the value of the scalar
field at the center of the soliton shows a rapid increase for the
same reason. The same discussion holds for every type of q-star
with differences that depend on the parameters of the Lagrangian.

The role of the frequency is also important. Small $\omega$ means
small value for the metric $A$ at the surface, consequently
stronger gravity, and a larger soliton to generate this gravity
in the case of flat spacetime
\cite{qstars-global,qstars-charged,qstars-nonabelian,qstars-fermion}.
So, a larger soliton in flat spacetime is \textit{too} large to
deserve its magnitude when $\Lambda\neq0$ and more sensitive in
the influence of the increase in absolute values of the
cosmological constant. That is why the consequent decrease in the
soliton parameters (radius, mass and charge) is more rapid for a
soliton with small $\omega$, i.e. for a \textit{large} soliton.

It will be interesting to compare our results with the
corresponding ones obtained by Astefanesei and Radu,
\cite{astefanesei}, when investigating a bosonic \textit{but not
solitonic} star within an AdS spacetime. They found that, in
general, the energy and charge for a boson star decrease with the
increase of the cosmological constant, but $\phi(0)$ increases.
These results are in agreement with what we obtained here (though
there is no a priori need for any agreement between two
substantially different field configurations) with the exception
of the initial increase in the values of $M$ and $Q$ in the
region of the small cosmological constant.

\initiate
\section{Q-stars with two scalar fields}

We will now describe a first generalization of the above simple
model, namely, a complex, scalar, N-carrying field $\phi$ and a
real one, $\sigma$, used to produce the special q-ball type
potential and to attribute mass to the N-carrying field. The
Lagrangian density is:
\begin{equation}\label{3.1}
\mathcal{L}/\sqrt{-g}=g^{\mu\nu}{(\partial_{\mu}\phi)}^{\ast}
(\partial_{\nu}\phi)+\frac{1}{2}g^{\mu\nu}(\partial_{\mu}\sigma)
(\partial_{\nu}\sigma)-U(\phi,\sigma)\ .
\end{equation}
In order to have static metric, we regard the real $\sigma$ time
independent and allow an harmonic time dependence to the $\phi$
field, writing:
\begin{equation}\label{3.2}
\phi(\vec{\rho},t)=\varphi(\rho)e^{-\imath\omega t}\ .
\end{equation}
In this case, Einstein equations take the form:
\begin{equation}\label{3.3}
\frac{e^{-\lambda}-1}{{\rho}^2}-e^{-\lambda}\frac{\lambda'}{\rho}=-8\pi
G\left[e^{-\nu}\omega^2\varphi^2+e^{-\lambda}{\left(\frac{d\varphi}{d\rho}\right)}^2+
e^{-\lambda}{\left(\frac{d\sigma}{d\rho}\right)}^2+U\right]-\Lambda\
,
\end{equation}
\begin{equation}\label{3.4}
\frac{e^{-\lambda}-1}{{\rho}^2}+e^{-\lambda}\frac{\nu'}{\rho}=8\pi
G\left[e^{-\nu}\omega^2\varphi^2+e^{-\lambda}{\left(\frac{d\varphi}{d\rho}\right)}^2+
e^{-\lambda}{\left(\frac{d\sigma}{d\rho}\right)}^2-U\right]-\Lambda\
,
\end{equation}
and the Lagrange equations:
\begin{equation}\label{3.5}
e^{-\lambda}\left\{\varphi''+[2/\rho+(1/2)(\nu'-\lambda')]\varphi'
\right\}+e^{-\nu}\omega^2\varphi-\frac{\partial
U(\varphi,\sigma)}{\partial\varphi^2}\varphi=0\ ,
\end{equation}
\begin{equation}\label{3.6}
e^{-\lambda}\left\{\sigma''+[2/\rho+(1/2)(\nu'-\lambda')]\sigma'
\right\}-\frac{\partial U(\varphi,\sigma)}{\partial\sigma}=0\ .
\end{equation}

\begin{figure}
\centering
\includegraphics{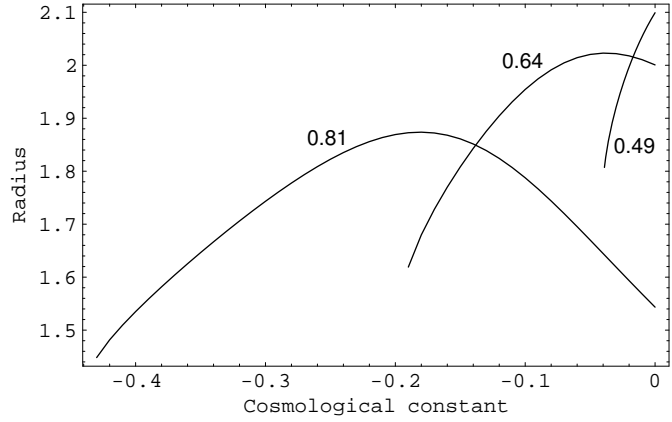}
\caption{The soliton radius for a q-star with two scalar fields as
a function of the cosmological constant for three different values
of $A_{\textrm{sur}}$.} \label{figure2.1}
\end{figure}

\begin{figure}
\centering
\includegraphics{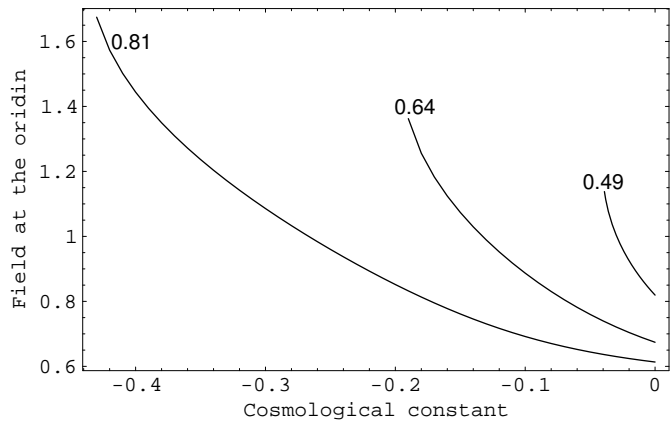}
\caption{The value of the N-carrying field at the center of a
q-star with two scalar fields as a function of the cosmological
constant for three different values of $A_{\textrm{sur}}$.}
\label{figure2.2}
\end{figure}

\begin{figure}
\centering
\includegraphics{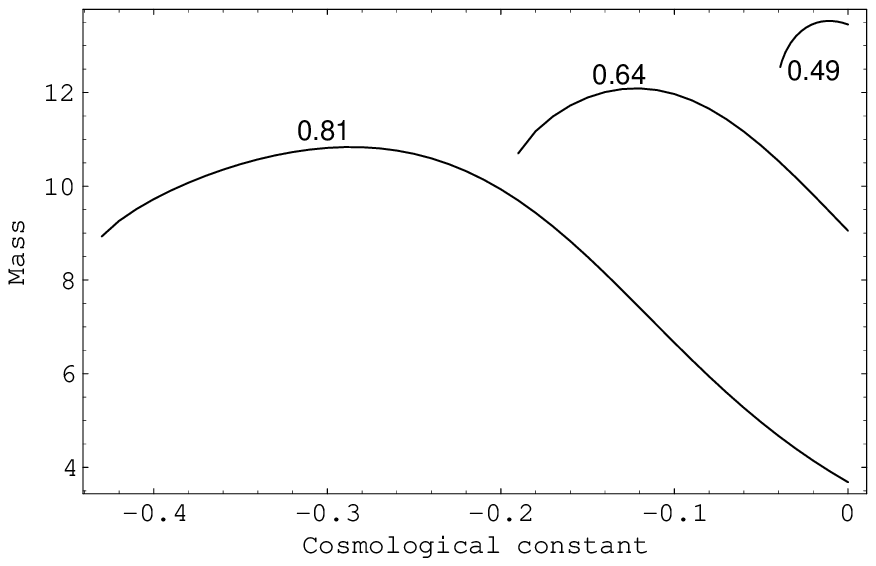}
\caption{The total mass for a q-star with two scalar fields as a
function of the cosmological constant for three different values
of $A_{\textrm{sur}}$.} \label{figure2.3}
\end{figure}

\begin{figure}
\centering
\includegraphics{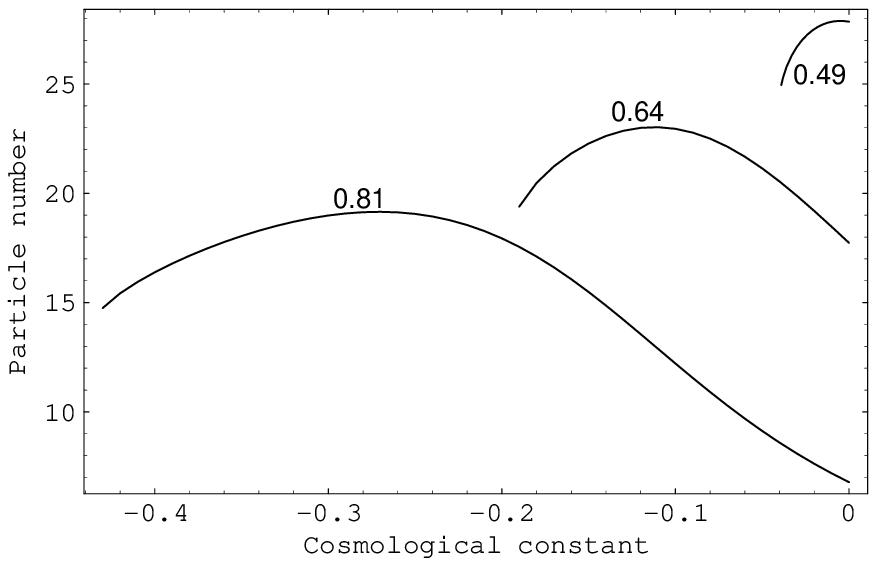}
\caption{The particle number for a q-star with two scalar fields
as a function of the cosmological constant for three different
values of $A_{\textrm{sur}}$.} \label{figure2.4}
\end{figure}

The potential has the following form:
\begin{equation}\label{3.7}
U=a|\phi|^2\sigma^2+b|\phi|^4+c(\sigma^2-d)^2\ .
\end{equation}
Defining $\mu$ and $\lambda$:
\begin{equation}\label{3.8}
\mu\equiv(2b)^{1/4}a^{-1/2}d^{-1/2}\ , \hspace{1em}
\lambda\equiv4cba^{-2}\ ,
\end{equation}
and rescaling the fields:
\begin{equation}\label{3.9}
\phi=\tilde{\phi}(2b)^{-1/4}\mu^{-1}\ , \hspace{1em}
\sigma=\tilde{\sigma}\mu^{-1}(2b)^{1/4}a^{-1/2}\ ,
\end{equation}
also rescaling $\rho$, $\omega$, $t$ and $\Lambda$ according to
eq. \ref{2.16} (replacing now $m$ with our new mass scale $\mu$),
we find that a $\mu^{-4}$ factor can be extracted from all energy
quantities. So, dropping the tildes and the $O(G\mu^2)$ terms we
find from the Euler-Lagrange equations that for the interior:
\begin{equation}\label{3.10}
\varphi^2=\omega^2B\ , \hspace{1em} \sigma\cong0\ ,
\end{equation}
\begin{equation}\label{3.11}
U=\frac{1}{2}(\omega^4B^2+\lambda), \hspace{1em} W=\omega^4B^2\ ,
\hspace{1em} V\cong0\ ,
\end{equation}
Einstein equations read:
\begin{equation}\label{3.12}
\frac{1-A}{r^2}-\frac{1}{r}\frac{dA}{dr}=\frac{3}{2}\omega^4B^2+\frac{1}{2}\lambda
+\Lambda\ ,
\end{equation}
\begin{equation}\label{3.13}
\frac{A-1}{r^2}-\frac{1}{r}\frac{A}{B}\frac{dB}{dr}=\frac{1}{2}\omega^4B^2-\frac{1}{2}\lambda
-\Lambda\ .
\end{equation}

The eigenvalue equation for the frequency results from a
discussion similar to that of eqs. \ref{2.28}-\ref{2.30}. We find
that:
\begin{equation}\label{3.14}
\omega=\lambda^{1/4}A_{\textrm{sur}}^{1/2}\ ,
\end{equation}
establishing in this way the frequency as a measure of the
gravity strength. We will choose $\lambda=1/9$. One can find that
the energy and charge of the solitonic configuration are:
\begin{equation}\label{3.15}
E=4\pi\int_0^Rdrr^2\left(\frac{3}{2}\omega^4B^2+\frac{1}{2}\lambda\right)\
,
\end{equation}
\begin{equation}\label{3.16}
Q=8\pi\int_0^Rdrr^2\omega\varphi^2\sqrt{\frac{B}{A}}=8\pi\int_0^Rdrr^2\omega^3B^{3/2}A^{-1/2}\
.
\end{equation}

In figures \ref{figure2.1}-\ref{figure2.4} we depict the main
results obtained by the numerical solutions of the Einstein
equations and the use of eqs. \ref{3.15}-\ref{3.16}. We reproduce
results similar to those obtained in q-stars with one scalar
field.

\initiate
\section{Non-abelian q-stars}

We choose for simplicity a field $\phi$ in the $SO(3)$
$\mathbf{5}$ representation. The Lagrangian is:
\begin{equation}\label{4.1}
\mathcal{L}=\frac{1}{2}\textrm{Tr}g^{\mu\nu}(\partial_{\mu}\phi)(\partial_{\nu}\phi)
-\textrm{Tr}U\ ,
\end{equation}
with $U$ a general renormalizable potential:
\begin{equation}\label{4.2}
U=\frac{\mu^2}{2}\phi^2+\frac{g}{3!}\phi^3+\frac{\lambda}{4!}\phi^4\
.
\end{equation}
Einstein equations read:
\begin{equation}\label{4.3}
\frac{e^{-\lambda}-1}{{\rho}^2}-e^{-\lambda}\frac{\lambda'}{\rho}=-8\pi
G\left[U+\textrm{Tr}\frac{1}{2}e^{-\nu}\left(\frac{\partial\phi}{\partial
t}\right)^2
+\textrm{Tr}\frac{1}{2}\left(\frac{\partial\phi}{\partial\rho}\right)^2\right]-\Lambda\
,
\end{equation}
\begin{equation}\label{4.4}
\frac{e^{-\lambda}-1}{{\rho}^2}+e^{-\lambda}\frac{\nu'}{\rho}=8\pi
G\left[\textrm{Tr}\frac{1}{2}e^{-\nu}\left(\frac{\partial\phi}{\partial
t}\right)^2
+\textrm{Tr}\frac{1}{2}\left(\frac{\partial\phi}{\partial\rho}\right)^2-U\right]-\Lambda\
.
\end{equation}
To simplify the above equations, we define:
\begin{equation}\label{4.5}
g=\mu\tilde{g}\ , \hspace{1em} \phi=(\mu/\tilde{g})\tilde{\phi}\ ,
\hspace{1em} \lambda=\tilde{g}^2\tilde{\lambda}\ , \hspace{1em}
\rho=\tilde{\rho}\mu^{-1}\ ,
\end{equation}
so as to extract a $\mu^4/\tilde{g}^2$ factor form every energy
quantity. The potential takes the simple form:
$$U=\frac{\mu^4}{\tilde{g}^2}\widetilde{U}\ , \hspace{1em}
\widetilde{U}=\frac{\tilde{\phi}^2}{2}+\frac{\tilde{\phi}^3}{3!}+\frac{\tilde{\lambda}}{4!}
\tilde{\phi}^4\ .$$ We also define:
\begin{equation}\label{4.6}
\tilde{r}=\tilde{\rho}\sqrt{8\pi G\frac{\mu^2}{\tilde{g}^2}}\ ,
\hspace{1em} 8\pi
G\frac{\mu^4}{\tilde{g}^2}\widetilde{\Lambda}=\Lambda\ .
\end{equation}

\begin{figure}
\centering
\includegraphics{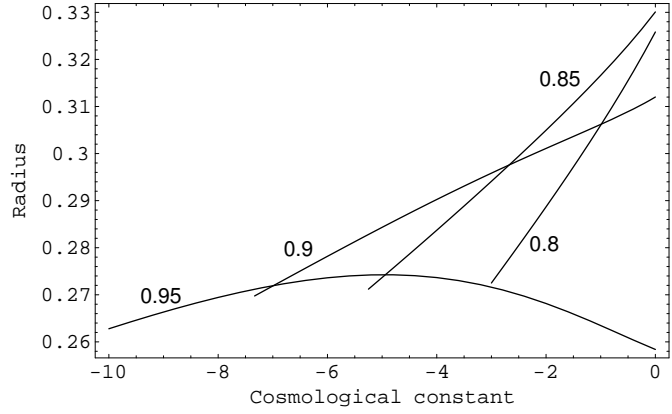}
\caption{The radius of a non-abelian q-star as a function of the
cosmological constant for four values of
$A_{\textrm{sur}}^{1/2}$.} \label{figure3.1}
\end{figure}

\begin{figure}
\centering
\includegraphics{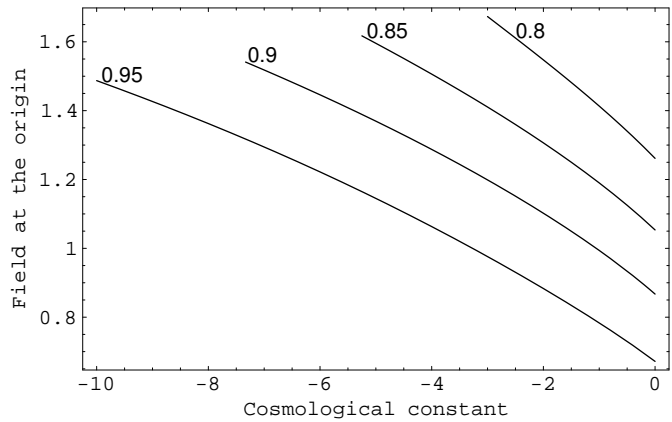}
\caption{The absolute value of the scalar field $\tilde{\phi}_2$
at the center of the a non-abelian q-star as a function of the
cosmological constant for four values of
$A_{\textrm{sur}}^{1/2}$.} \label{figure3.2}
\end{figure}

\begin{figure}
\centering
\includegraphics{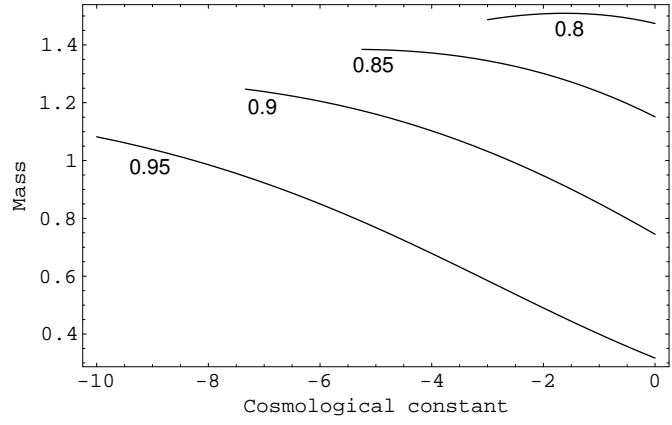}
\caption{The total mass for a non-abelian q-star as a function of
the cosmological constant for four values of
$A_{\textrm{sur}}^{1/2}$.} \label{figure3.3}
\end{figure}

\begin{figure}
\centering
\includegraphics{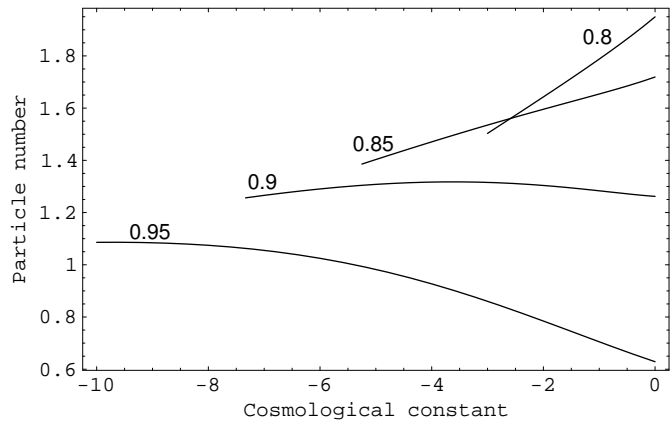}
\caption{The particle number for a non-abelian q-star as a
function of the cosmological constant for four values of
$A_{\textrm{sur}}^{1/2}$.} \label{figure3.4}
\end{figure}

With the above redefinitions, Einstein equations take the form:
\begin{equation}\label{4.7}
\frac{1-A}{\tilde{r}^2}-\frac{1}{\tilde{r}}\frac{dA}{d\tilde{r}}=
\mathcal{U}+\mathcal{W}+\mathcal{V}+\widetilde{\Lambda}\ ,
\end{equation}
\begin{equation}\label{4.8}
\frac{A-1}{\tilde{r}^2}-\frac{1}{\tilde{r}}\frac{A}{B}\frac{dB}{d\tilde{r}}=
\mathcal{W}+\mathcal{V}-\mathcal{U}-\widetilde{\Lambda}\ ,
\end{equation}
where:
\begin{equation}\label{4.9}
\begin{split}
\mathcal{U}&=\textrm{Tr}\widetilde{U}\ , \\
\mathcal{W}&=\textrm{Tr}\frac{1}{2}B\left(\frac{\partial\tilde{\phi}}{\partial\tilde{t}}
\right)^2\ , \\ \mathcal{V}&=\textrm{Tr}\frac{1}{2}A
\left(\frac{\partial\tilde{\phi}}{\partial\tilde{r}}\right)^28\pi
G\frac{\mu^2}{\tilde{g}^2}\ .
\end{split}
\end{equation}
We define:
\begin{equation}\label{4.10}
\frac{\partial\phi}{\partial t}=\imath[\Omega,t]\ , \hspace{1em}
\Omega\equiv\mu\widetilde{\Omega}\equiv
-\imath\tilde{\omega}\mu\left(
\begin{array}{ccc}
  0 & 0 & 1 \\
  0 & 0 & 0 \\
  -1 & 0 & 0
\end{array}\right)\ .
\end{equation}
The Lagrange equation reads:
\begin{eqnarray}\label{4.11}
-B[\widetilde{\Omega},[\widetilde{\Omega},\tilde{\phi}]]+
\frac{\partial\mathcal{U}}{\partial\tilde{\phi}}-
\frac{1}{3}\mathbf{1}\textrm{Tr}\left(\frac{\partial\mathcal{U}}{\partial\tilde{\phi}}\right)=
\nonumber\\ 8\pi
G\frac{{\mu}^2}{{\tilde{g}}^2}\left[\frac{{\partial}^2\tilde{\phi}}{\partial{\tilde{r}}^2}
+\frac{\partial\tilde{\phi}}{\partial{\tilde{r}}}\left(\frac{2}{\tilde{r}}+
\frac{1}{2A}\frac{dA}{d\tilde{r}}-\frac{1}{2B}\frac{dB}{d\tilde{r}}\right)\right]\
.
\end{eqnarray}

We diagonalize $\phi=e^{\imath R}\phi_{\textrm{diag}}e^{-\imath
R}$. The rigid rotation condition implies that: $R(\rho,t)=\Omega
t +C$, where the constant $C$ is eliminated through a global
$SO(3)$ rotation. Diagonalizing, we finally take
\begin{equation}\label{4.12}
\tilde{\phi}=-\frac{1}{2}\tilde{\phi}_2\cdot\textrm{diag}(1+y,-2,1-y)\
.
\end{equation}
Inserting eq. \ref{4.12} into \ref{4.11} and dropping the
$O(G\mu^2/\tilde{g}^2)$ terms we find:
\begin{equation}\label{4.13}
\tilde{\phi}+\frac{1}{2}\tilde{\phi}^2+\frac{1}{6}\tilde{\lambda}\tilde{\phi}^3-
\frac{1}{3}\mathbf{1}\textrm{Tr}\left(\frac{1}{2}\tilde{\phi}^2+
\frac{1}{6}\tilde{\phi}^3\right)=
2\tilde{\omega}^2B(\tilde{\phi}_1-\tilde{\phi}_3)\textrm{diag}(1,0,-1)\
,
\end{equation}
where $\tilde{\phi}_{1,3}=\tilde{\phi}_2(1\pm y)$. Taking the
$(2,2)$ component of the above equation, we find:
\begin{equation}\label{4.14}
y^2(\tilde{r})=\frac{3}{\tilde{\phi}_2}
\frac{\tilde{\lambda}\tilde{\phi}_2^2+2\tilde{\phi}_2+8}{2-\tilde{\lambda}\tilde{\phi}_2}\
.
\end{equation}
The above equation, combined with the $2\mathcal{W}=\textrm{Tr}
(\tilde{\phi}\cdot\partial\mathcal{U}/\partial\tilde{\phi})$
relation (resulting from the Euler-Lagrange \ref{4.11}, when
dropping $O(G\mu^2/\tilde{g}^2)$ terms, multiplying by
$\tilde{\phi}$ and then tracing) gives for the $\tilde{\phi}_2$
field:
\begin{equation}\label{4.15}
\tilde{\phi}_2(\tilde{r})=
\frac{1-4\tilde{\lambda}\tilde{\omega}^2B(\tilde{r})}{2\tilde{\lambda}}+
\frac{1}{2\tilde{\lambda}}
\{[1-4\tilde{\lambda}\tilde{\omega}^2B(\tilde{r})]^2+
8\tilde{\lambda}[4\tilde{\omega}^2B(\tilde{r})-1]\}^{1/2}\ .
\end{equation}
We will now find an eigenvalue equation for $\tilde{\omega}$. At
the surface holds:
\begin{equation}\label{4.16}
(\mathcal{U}-\mathcal{W}-\mathcal{V})_{\textrm{sur}}=0\ .
\end{equation}
If $R$ the star radius, using eq. \ref{4.14} (which at the inner
edge of the surface remains valid, because the spatial derivative
of the matter field is of $O(G\mu^2/\tilde{g}^2)^{1/2}$ at this
edge) and eq. \ref{4.16} we find:
\begin{equation}\label{4.17}
\tilde{\phi}_2(R)=\frac{12}{\tilde{\lambda}}
\left[\frac{y^2-1}{(y^2+3)^2}\right]_{\textrm{sur}}\ ,
\end{equation}
\begin{equation}\label{4.18}
{\tilde{\omega}}^2B(R)={\left[\frac{1}{4}\left(1+\frac{3}{y^2}\right)-
\frac{3}{4\tilde{\lambda}}\frac{1}{y^2}
\frac{{(y^2-1)}^2}{{(y^2+3)}^2}\right]}_{\textrm{sur}}\ .
\end{equation}
Solving now \ref{4.14} (which holds true when $\mathcal{V}\cong0$)
with respect to $\tilde{\phi}_2$ and substituting in eq.
\ref{4.17} we find:
\begin{equation}\label{4.19}
{[(\tilde{\lambda}-1){(y^2+3)}^3+16{(y^2+3)}^2-72(y^2+3)+96]}_{\textrm{sur}}=0\
.
\end{equation}
The solution to the above equation is:
\begin{eqnarray}\label{4.20}
y^2(R)=-3+\frac{16}{3(1-\tilde{\lambda})}+ \nonumber\\
\frac{2}{3}\frac{2^{1/3}}{-1+\tilde{\lambda}}
{\left[-13+162\tilde{\lambda}-81{\tilde{\lambda}}^2+9{(-1+\tilde{\lambda})}^2
\sqrt{\frac{-1+81{\tilde{\lambda}}^2}{{(-1+\tilde{\lambda})}^2}}\right]}^{1/3}+
\nonumber\\ +\frac{2}{3}\frac{2^{1/3}}{1-\tilde{\lambda}}
{\left[13+162\tilde{\lambda}+81{\tilde{\lambda}}^2+9{(-1+\tilde{\lambda})}^2
\sqrt{\frac{-1+81{\tilde{\lambda}}^2}{{(-1+\tilde{\lambda})}^2}}\right]}^{1/3}\
.
\end{eqnarray}
So, we give a certain value to the $\tilde{\lambda}$ parameter,
say $\tilde{\lambda}=2/3$, and we find the value of $y^2(R)$, from
eq. \ref{4.20}. We then substitute that value into eq. \ref{4.18}
(finding that $\tilde{\omega}=0.4956A_{\textrm{sur}}^{1/2}$) and
going back to \ref{4.15} and \ref{4.14} we insert the fields in
Einstein equations and solve them numerically. The total energy
of the field configuration is computed with the help of $T^0_{\
0}$ component of the energy-momentum tensor, equal to
$\mathcal{W}+\mathcal{U}$, and the total charge is defined as:
\begin{equation}\label{4.21}
Q=-\imath\int
d^3\tilde{r}\sqrt{|g|}B[\tilde{\phi},\dot{\tilde{\phi}}]=4\pi\imath\tilde{g}
\int_0^Rd\tilde{r}{\tilde{r}}^2\sqrt{B/A}{\tilde{\phi}}^2_2y^2\tilde{\omega}
\left(\begin{array}{ccc}
  0 & 0 & 1 \\
  0 & 0 & 0 \\
  -1 & 0 & 0
\end{array}\right) \ .
\end{equation}
For the calculation of the charge we use the quantity:
$4\pi\int_0^Rd\tilde{r}{\tilde{r}}^2\sqrt{B/A}{\tilde{\phi}}^2_2y^2\tilde{\omega}$.

\initiate
\section{Scalar-fermion q-stars}

We now consider a realistic field configuration  composed of one
real scalar field and one fermionic. The fermion carries the
particle number, and its mass is generated by the interaction with
the scalar field. Within the soliton, the scalar field $\sigma$
is approximately zero and the fermion mass zero, but outside the
soliton, the fermion has a certain mass, the scalar has an
approximately constant, $\sigma_0$, value, eliminating in this
way any potential energy in a potential of
$(\sigma^2-\sigma_0^2)^2$ type. So, the question for the stability
of a fermion-scalar soliton depends, in a few words, on the
energy difference of a $\sigma_0^4$-type potential energy from the
soliton interior plus the consequent kinetic energy of the trapped
fermions, minus total the mass of the free fermions with the same
charge. Because the fermion mass, supposed to result from a
$g\sigma\bar{\psi}\psi$ type of interaction, can be taken as
large as we want, theoretically and by changing $g$, the soliton
stability with respect to decay into free particles is always
granted. Here, we will choose a $g$ of the same order of magnitude
as the other Lagrangian parameters.

Regarding the fermion scalar q-star as a zero temperature,
spherically symmetric fermionic sea with local Fermi energy and
momentum with: $\varepsilon_F^2=k_F^2+m^2(\sigma)$, we have for
the local and scalar fermion density respectively:
\begin{equation}\label{5.1}
\langle\psi^{\dagger}\psi\rangle=\frac{2}{8\pi^3}\int
n_kd^3k=\frac{k_F^3}{3\pi^2}\ ,
\end{equation}
\begin{equation}\label{5.2}
\langle\bar{\psi}\psi\rangle=\frac{2}{8\pi^3}\int n_kd^3k\frac{m}
{(k^2+m^2)^{1/2}}=\frac{m}{2\pi^2}
\left(k_F\varepsilon_F-m^2\ln\frac{k_F+\varepsilon_F}{m}\right)\ .
\end{equation}
Fermion energy and pressure density are:
\begin{equation}\label{5.3}
P_{\psi}=\frac{2}{8\pi^3}\int n_kd^3k\frac{k^2}{3(k^2+m^2)^{1/2}}=
\frac{1}{4}(\varepsilon_F\langle\psi^{\dagger}\psi\rangle-m\langle\bar{\psi}\psi\rangle)\
,
\end{equation}
\begin{equation}\label{5.4}
\mathcal{E}_{\psi}=\frac{2}{8\pi^2}\int
n_kd^3k(k^2+m^2)^{1/2}=3P_{\psi}+m\langle\bar{\psi}\psi\rangle\ .
\end{equation}
When including gravity, the Lagrangian density is:
\begin{equation}\label{5.5}
\mathcal{L}/\sqrt{-g}=\frac{\imath}{2}(\bar{\psi}{\gamma}^{\mu}{\psi}_{;\mu}
-\bar{\psi}_{;\mu}{\gamma}^{\mu}\psi)
-m(\sigma)\bar{\psi}\psi+\frac{1}{2}{\sigma}_{;\mu}{\sigma}^{;\mu}-U(\sigma)\
,
\end{equation}
with
\begin{equation}\label{5.6}
m(\sigma)=g\sigma\ ,
\end{equation}
(we will choose g=10 for our calculations) and:
\begin{equation}\label{5.7}
U=\frac{1}{4}\lambda{({\sigma}^2-{\sigma}_0^2)}^2\ .
\end{equation}
The equation of motion for the scalar field is:
\begin{equation}\label{5.8}
A\{\sigma''+[2/\rho+(1/2)(A'/A-B'/B)]\sigma'\}=-\frac{\partial}{\partial\sigma}(P_{\psi}-U)\
.
\end{equation}
We rescale with respect to the mass scale $\sigma_0$ as follows:
\begin{equation}\label{5.9}
\tilde{\sigma}=\sigma/\sigma_0\ , \hspace{1em}
\tilde{\psi}=\psi/\sigma_0^{3/2}\ , \hspace{1em}
\tilde{m}=m/\sigma_0\ , \hspace{1em} \tilde{\rho}=\rho\sigma_0 \,
\hspace{1em} \widetilde{\mathcal{L}}=\mathcal{L}/\sigma^4_0\ ,
\end{equation}
and define:
\begin{equation}\label{5.10}
\tilde{r}=\tilde{\rho}\sqrt{8\pi G\sigma_0^2}\ , \hspace{1em}
\widetilde{\Lambda}8\pi G\sigma_0^4=\Lambda\ .
\end{equation}

In the presence of gravity, we replace the Fermi energy with a
global chemical potential, $\omega_{\psi}$ , so as to make the
theory generally covariant, by establishing $\varepsilon_F$ as
the zero component of a four vector, \cite{fr3}. So we define:
\begin{equation}\label{5.11}
{\omega}_{\psi}^2={\varepsilon}_F^2B^{-1}(r)\ .
\end{equation}
The energy-momentum tensor is:
\begin{equation}\label{5.12}
T^{\mu\nu}=\frac{1}{2}\imath(\bar{\psi}{\gamma}^{(\mu}{\psi}^{;\nu)}-
{\bar{\psi}}^{;(\mu}{\gamma}^{\nu)}\psi)+
{\sigma}^{;\mu}{\sigma}^{;\nu}-g^{\mu\nu}[1/2{\sigma}_{;\alpha}{\sigma}^{;\alpha}-U(\sigma)]\
,
\end{equation}
so, Einstein equations read, after rescaling, dropping the tildes
and ignoring the $O(G\sigma_0^2)$ terms:
\begin{equation}\label{5.13}
\frac{1-A}{r^2}-\frac{1}{r}\frac{dA}{dr}=\mathcal{E}_{\psi}+U+\Lambda\
,
\end{equation}
\begin{equation}\label{5.14}
\frac{A-1}{r^2}-\frac{1}{r}\frac{A}{B}\frac{dB}{dr}=P_{\psi}-U-\Lambda\
.
\end{equation}
There is one unknown quantity, namely the chemical potential,
playing the role of ``eigen"-frequency here. Within the surface
holds:
\begin{equation}\label{5.15}
U=P_{\psi}\Rightarrow\frac{\lambda}{4}=\frac{1}{12\pi^2}{\omega}^4_{\psi}B^2_{\textrm{sur}}\
,
\end{equation}
so Einstein equations are:
\begin{equation}\label{5.16}
\frac{1-A}{r^2}-\frac{1}{r}\frac{dA}{dr}=\frac{1}{4\pi}{\omega}^4_{\psi}B^2
+\frac{\lambda}{4}+\Lambda\ ,
\end{equation}
\begin{equation}\label{5.17}
\frac{A-1}{r^2}-\frac{1}{r}\frac{A}{B}\frac{dB}{dr}=\frac{1}{12\pi}{\omega}^4_{\psi}B^2
-\frac{\lambda}{4}-\Lambda\ .
\end{equation}
We choose $\lambda=1/2$. The particle number is given from the
relation:
\begin{equation}\label{5.18}
Q_{\psi}=\int\sqrt{-g}d^3xj^0=4\pi\int
drr^2\sqrt{\frac{1}{A}}\langle\psi^{\dagger}\psi\rangle=
\frac{4}{3\pi}\int drr^2\omega_{\psi}^3B^{3/2}A^{-1/2}\ .
\end{equation}

\begin{figure}
\centering
\includegraphics{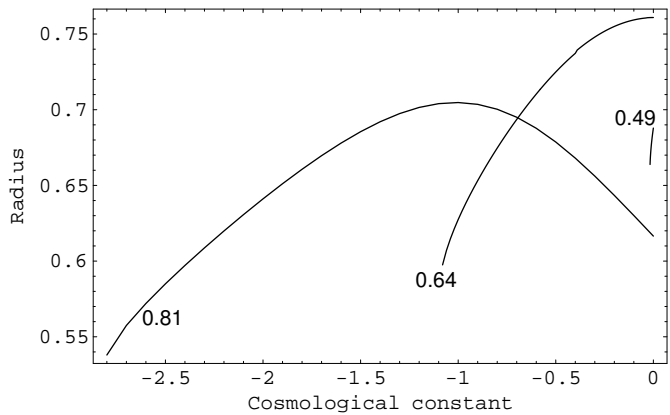}
\caption{The radius of a fermion-scalar q-star as a function of
the cosmological constant for three different values of
$A_{\textrm{sur}}$.} \label{figure4.1}
\end{figure}

\begin{figure}
\centering
\includegraphics{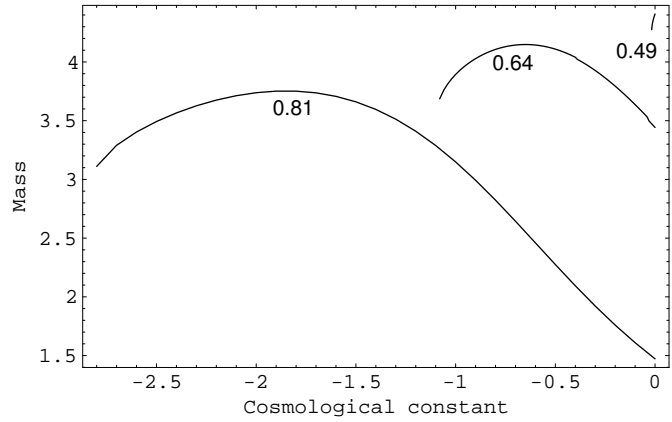}
\caption{The total mass for a fermion-scalar q-star as a function
of the cosmological constant for three different values of
$A_{\textrm{sur}}$.} \label{figure4.2}
\end{figure}

\begin{figure}
\centering
\includegraphics{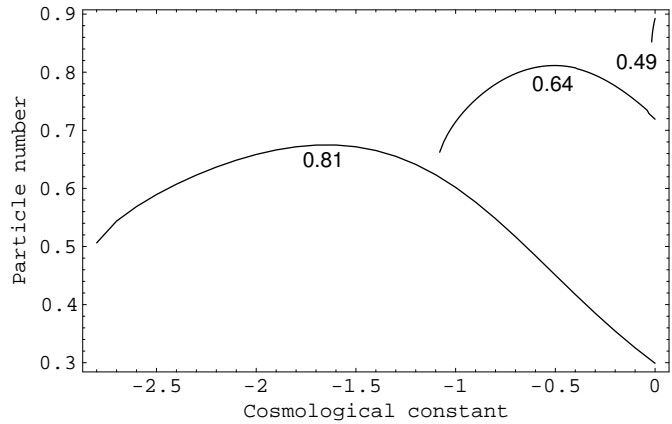}
\caption{The particle number for a fermion-scalar q-star as a
function of the cosmological constant for three different values
of $A_{\textrm{sur}}$.} \label{figure4.3}
\end{figure}

\initiate
\section{Charged q-stars}

\textbf{Charged q-stars with one scalar field}

\vspace{1em}

The action for a scalar field with a $U(1)$ local symmetry
coupled to gravity is:
\begin{equation}\label{6.1}
S=\int d^4x\sqrt{-g}\left[\frac{R-2\Lambda}{16\pi
G}+g^{\mu\nu}{(D_{\mu}\phi)}^{\ast}(D_{\nu}\phi)-U-\frac{1}{4}F_{\mu\nu}F^{\mu\nu}\right]
\end{equation}
with:
$$F_{\mu\nu}\equiv{\partial}_{\mu}A_{\nu}-{\partial}_{\nu}A_{\mu}\ ,
\hspace{1em} D_{\mu}\phi\equiv{\partial}_{\mu}\phi-\imath
eA_{\mu}\phi\ ,$$ with $e$ the charge or field strength. The total
charge $Q$ will be hereafter called particle number, identified
with the total electric charge when the charge $e$ of each
individual particle is unity. The energy-momentum tensor is:
\begin{eqnarray}\label{6.2}
T_{\mu\nu}={(D_{\mu}\phi)}^{\ast}(D_{\mu}\phi)+
(D_{\mu}\phi){(D_{\mu}\phi)}^{\ast}-
g_{\mu\nu}[g^{\alpha\beta}{(D_{\alpha}\phi)}^{\ast}(D_{\beta}\phi)]
\nonumber\\
-g_{\mu\nu}U-\frac{1}{4}g_{\mu\nu}F^{\alpha\beta}F_{\alpha\beta}+g^{\alpha\beta}
F_{\nu\alpha}F_{\nu\beta}\ .
\end{eqnarray}
The Euler-Lagrange equation for the scalar field is:
\begin{equation}\label{6.3}
\left[\frac{1}{\sqrt{-g}}D_{\mu}(\sqrt{-g}g^{\mu\nu}D_{\mu})-\frac{dU}{d|\phi|^2}\right]\phi\
.
\end{equation}
The Noether current is:
\begin{eqnarray}\label{6.4}
j^{\mu}=\sqrt{-g}g^{\mu\nu}\imath({\phi}^{\ast}D_{\nu}\phi- \phi
D_{\nu}{\phi}^{\ast})= \nonumber\\
\sqrt{-g}g^{\mu\nu}({\phi}^{\ast}{\partial}_{\nu}\phi-
\phi{\partial}_{\nu}{\phi}^{\ast} +2e^2A_{\nu}{|\phi|}^2)\ .
\end{eqnarray}
The Noether charge resulting form the above current is conserved.

We use the same ansatz for the scalar field, namely the
$\phi=\sigma(r)e^{-\imath\omega t}$, and we assume no magnetic
fields choosing $A_{\mu}=(A_0,0,0,0)$. We are now be able to use
the same static metric. With these assumptions, Einstein equations
read:
\begin{eqnarray}\label{6.5}
\frac{e^{-\lambda}-1}{{\rho}^2}-e^{-\lambda}\frac{\lambda'}{\rho}=
\nonumber\\ -8\pi
G\left[{(\omega+eA_0)}^2e^{-\nu}{\sigma}^2+U+{\sigma'}^2e^{-\lambda}+
(1/2){A'}^2_0e^{-\nu-\lambda} \right]-\Lambda\ ,
\end{eqnarray}
\begin{eqnarray}\label{6.6}
\frac{e^{-\lambda}-1}{{\rho}^2}+e^{-\lambda}\frac{\nu'}{\rho}=
\nonumber\\ 8\pi
G\left[{(\omega+eA_0)}^2e^{-\nu}{\sigma}^2-U+{\sigma'}^2e^{-\lambda}-
(1/2){A'}^2_0e^{-\nu-\lambda} \right]-\Lambda\ ,
\end{eqnarray}
and the Lagrange equations for the scalar and gauge field are
respectively:
\begin{equation}\label{6.7}
\sigma''+\left[\frac{2}{\rho}+\frac{1}{2}(\nu'-\lambda')\right]\sigma'
+e^{\lambda}{(\omega+eA_0)}^2e^{-\nu}\sigma-e^{\lambda}\frac{dU}{d{\sigma}^2}\sigma=0
\ ,
\end{equation}
\begin{equation}\label{6.8}
A_0''+\left[\frac{2}{\rho}-\frac{1}{2}(\nu'+\lambda')\right]A_0'-2e{\sigma}^2
e^{\lambda}(\omega+eA_0)=0\ .
\end{equation}

Because, due to the covariant derivative, the $\omega+eA_0$
combination is very common, we define:
\begin{equation}\label{6.9}
\theta\equiv\omega+eA_0\ .
\end{equation}
We use eqs. \ref{2.15}-\ref{2.17} and rescale $\theta$ in the same
way as $\omega$. The rescaling for the field strength is similar
to the one used in charged boson (but not soliton) stars:
\begin{equation}\label{6.10}
\tilde{e}=e{\epsilon}^{-1}\ ,
\end{equation}
Einstein equations read, dropping the tildes and the $O(\epsilon)$
terms:
\begin{equation}\label{6.11}
\frac{1-A}{r^2}-\frac{1}{r}\frac{dA}{dr}={\theta}^2{\sigma}^2B+U+\frac{1}{2e^2}
{\left(\frac{d\theta}{dr}\right)}^2AB+\Lambda\ ,
\end{equation}
\begin{equation}\label{6.12}
\frac{A-1}{r^2}-\frac{1}{r}\frac{A}{B}\frac{dB}{dr}={\theta}^2{\sigma}^2B-U-\frac{1}{2e^2}
{\left(\frac{d\theta}{dr}\right)}^2AB-\Lambda\ ,
\end{equation}
and the Lagrange equations are:
\begin{equation}\label{6.13}
\theta^2B-\frac{dU}{d\sigma^2}=0\ ,
\end{equation}
\begin{equation}\label{6.14}
\frac{d^2\theta}{dr^2}+\left[\frac{2}{r}+\frac{1}{2}
\left(\frac{1}{A}\frac{dA}{dr}+\frac{1}{B}\frac{dB}{dr}\right)\right]\frac{d\theta}{dr}-
\frac{2e^2\sigma^2\theta}{A}=0\ ,
\end{equation}
with initial conditions $A(0)=1$, $B(\infty)=1$ and $A_0'(0)=0$.
The other initial condition for the value of the gauge field can
be received by repeating the discussion referring to the surface,
because the frequency $\omega$ is now replaced by the dynamical
variable $\theta$. So, eq. \ref{2.30} is replaced by:
\begin{equation}\label{6.15}
\theta_{\textrm{sur}}=\frac{A_{\textrm{sur}}^{1/2}}{2}=\frac{B_{\textrm{sur}}^{-1/2}}{2}\
.
\end{equation}
Eq. \ref{6.13} gives:
\begin{equation}\label{6.16}
\sigma^2=1+\theta\sqrt{B}\ ,
\end{equation}
\begin{equation}\label{6.17}
U=\frac{1}{3}(1+\theta^3B^{/32})\ .
\end{equation}
The energy in the interior of the field configuration is:
\begin{equation}\label{6.18}
E_{\textrm{int}}=4\pi\int_0^Rdrr^2\left[\theta^2\sigma^2B+\frac{1}{3}(1+\theta^3B^{3/2})+
\frac{1}{2e^2}{\left(\frac{d\theta}{dr}\right)}^2AB\right]
\end{equation}
and the particle number is:
\begin{equation}\label{6.19}
Q=8\pi\int_0^Rdrr^2\theta\sigma^2\sqrt{B/A}\ .
\end{equation}
The total electric charge is $eQ$.

At the exterior the Einstein and Lagrange equations take the form:
\begin{equation}\label{6.20}
1-A-r\frac{dA}{dr}=r^2\left[\Lambda+\frac{1}{2e^2}{\left(\frac{d\theta}{dr}\right)}^2AB\right]\
,
\end{equation}
\begin{equation}\label{6.21}
A-1-r\frac{A}{B}\frac{dB}{dr}=-r^2\left[\Lambda+\frac{1}{2e^2}
{\left(\frac{d\theta}{dr}\right)}^2AB\right]\ ,
\end{equation}
\begin{equation}\label{6.22}
\frac{d^2\theta}{dr^2}+\left[\frac{2}{r}+\frac{1}{2}
\left(\frac{1}{A}\frac{dA}{dr}+\frac{1}{B}\frac{dB}{dr}\right)\right]\frac{d\theta}{dr}=0\
.
\end{equation}
The above equations can be solved analytically:
\begin{equation}\label{6.23}
\theta(r)=\frac{-\theta'_{\textrm{sur}}R^2+r(\theta'_{\textrm{sur}}R+\theta_{\textrm{sur}})}{r}\
,
\end{equation}
\begin{equation}\label{6.24}
\begin{split}
A(r)&=\frac{{\theta'}^2_{\textrm{sur}}R^3(R-r)}{2e^2r^2}+
\frac{3RA_{\textrm{sur}}+3(r-R)-\Lambda(r^3-R^3)}{3r}\ , \\
B(r)&=\frac{1}{A(r)}\ .
\end{split}
\end{equation}
The total energy from the exterior is:
\begin{equation}\label{6.25}
E_{\textrm{ext}}=\frac{2\pi R^3{\theta'}^2_{\textrm{sur}}}{e^2}\ .
\end{equation}
The same discussion referring to the exterior holds for every kind
of charged q-star. The total energy of the field configuration
can be calculated either by eqs. \ref{6.18}, \ref{6.25}, or by
the relation:
$$A=1-2GE_{\textrm{tot}}/\rho+G(eQ)^2/(4\pi\rho^2)-\Lambda\rho^2/3\ , \hspace{1em}
\rho\rightarrow\infty\ ,$$ together with eq. \ref{6.19}.

\begin{figure}
\centering
\includegraphics{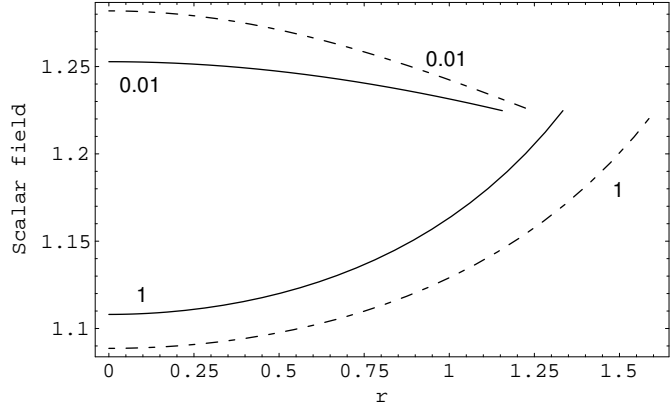}
\caption{The absolute values $|\sigma(r)|$ as a function of
radius, for $\Lambda=0$ (solid lines) and $\Lambda=-0.17$ (dashed
lines) for two different charges for a charged q-star with one
scalar field.} \label{figurec.01}
\end{figure}

\begin{figure}
\centering
\includegraphics{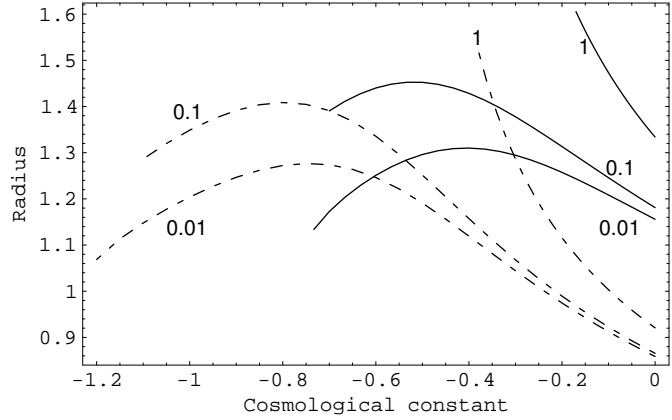}
\caption{The soliton radius as a function of the cosmological
constant for three different charges and for
$A_{\textrm{sur}}=0.64$ or, equivalently,
$\theta_{\textrm{sur}}=0.4$ (solid lines) and
$A_{\textrm{sur}}=0.81$ or, equivalently,
$\theta_{\textrm{sur}}=0.45$ (dashed lines), for a charged q-star
with one scalar field. Using the same values for the parameters we
produce figures \ref{figurec.2}-\ref{figurec.3}.}
\label{figurec.1}
\end{figure}

\begin{figure}
\centering
\includegraphics{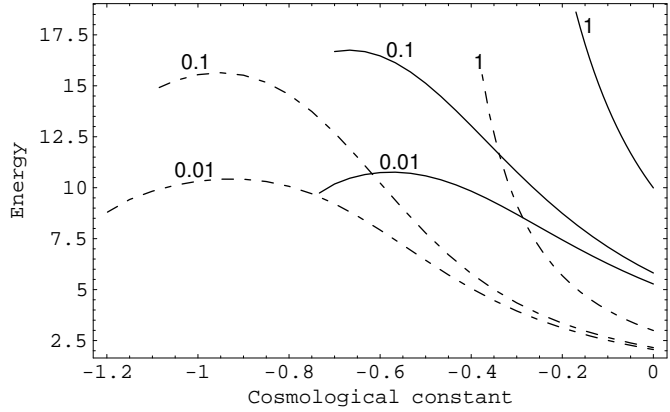}
\caption{The soliton mass as a function of the cosmological
constant for three different charges and two values of
$A_{\textrm{sur}}$, $0.64$ (solid lines) and $0.81$ (dashed
lines) for a charged q-star with one scalar field.}
\label{figurec.2}
\end{figure}

\begin{figure}
\centering
\includegraphics{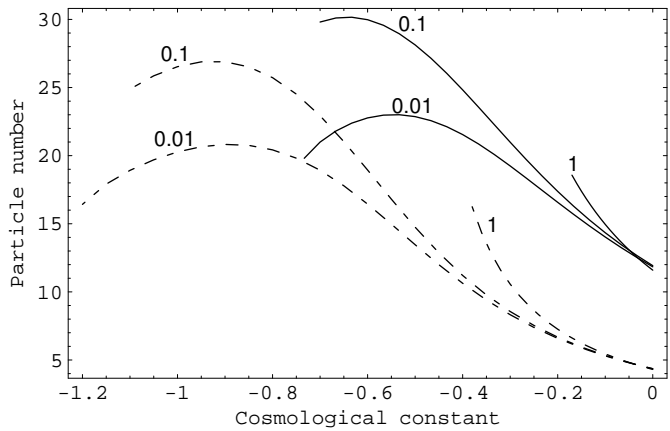}
\caption{The particle number of the soliton as a function of the
cosmological constant for three different charges and two values
of $A_{\textrm{sur}}$, $0.64$ (solid lines) and $0.81$ (dashed
lines) for a charged q-star with one scalar field.}
\label{figurec.3}
\end{figure}

From figures \ref{figurec.01}-\ref{figurec.3} we see that for
large charge the soliton radius, mass and particle number are in
general larger. For large values of the coupling $e$, here for
$e=1$, we stop the calculations when reaching the
$E_{\textrm{soliton}}=E_{\textrm{free}}$ area, where decay into
free particles is energetically favorable. Figure
\ref{figurec.01} has a very interesting interpretation. We see
that, for larger field strength, the soliton profile is larger
and the field tends to concentrate near the surface. Both of
these effects result from the electrostatic repulsion between the
different parts of the star.

\vspace{1em}

\textbf{Charged q-stars with two scalar fields}

\vspace{1em}

The Lagrangian density for a charged scalar field coupled to a
real one and to gravity is:
\begin{equation}\label{6.26}
\mathcal{L}/\sqrt{-g}=g^{\mu\nu}(D_{\mu}\phi)^{\ast}(D_{\nu}\phi)+
\frac{1}{2}g^{\mu\nu}(\partial_{\mu}\sigma)(\partial_{\nu}\sigma)-U-
\frac{1}{4}F^{\mu\nu}F_{\mu\nu}\ .
\end{equation}
We make the same rescalilngs for the fields, and use the ansatz of
eq. \ref{3.2} and the rescaled potential of eq. \ref{3.7}. Using
the Lagrange equation for the charged field we find:
\begin{equation}\label{6.27}
\varphi^2=\theta^2B\ , \hspace{1em}
U=\frac{1}{2}(\theta^4B^2+\lambda)\ , \hspace{1em} W=\theta^4B^2\
.
\end{equation}
Repeating the discussion that led to eq. \ref{6.15} we find the
value of the gauge field at the surface:
\begin{equation}\label{6.28}
\theta_{\textrm{sur}}=\left(\frac{\lambda}{B_{\textrm{sur}}^2}\right)^{1/4}\
.
\end{equation}
We finally have to solve numerically a system of coupled
equations governing the soliton interior: The two independent
Einstein equations, namely:
\begin{equation}\label{6.29}
\frac{1-A}{r^2}-\frac{1}{r}\frac{dA}{dr}=\frac{3}{2}\theta^4B^2+\frac{\lambda}{2}
+\frac{1}{2e^2}\left(\frac{d\theta}{dr}\right)^2+\Lambda\ ,
\end{equation}
\begin{equation}\label{6.30}
\frac{A-1}{r^2}-\frac{1}{r}\frac{A}{B}\frac{dB}{dr}=\frac{1}{2}\theta^4B^2-\frac{\lambda}{2}
-\frac{1}{2e^2}\left(\frac{d\theta}{dr}\right)^2-\Lambda
\end{equation}
and the equation of motion for the gauge field:
\begin{equation}\label{6.31}
\frac{d^2\theta}{dr^2}+\left[\frac{2}{r}+\frac{1}{2}
\left(\frac{1}{A}\frac{dA}{dr}+\frac{1}{B}\frac{dB}{dr}\right)\right]\frac{d\theta}{dr}-
\frac{2e^2\varphi^2\theta}{A}=0\ .
\end{equation}
The particle number is given by:
\begin{equation}\label{6.32}
Q=8\pi\int_0^R drr^2\theta^3B^{3/2}A^{-1/2}\ .
\end{equation}

\vspace{1em}

\textbf{Charged fermion-scalar q-stars}

\vspace{1em}

We now study a more realistic model, composed of a charged
fermion and a neutral scalar field, taken to be real for
simplicity. The Lagrangian density for such a system coupled to
gravity is:
\begin{equation}\label{6.33}
\mathcal{L}/\sqrt{-g}=\frac{\imath}{2}[\bar{\psi}\gamma^{\mu}(D_{\mu}\psi)-
(D_{\mu}\bar{\psi})\gamma^{\mu}\psi]-g\sigma\bar{\psi}\psi
+\frac{1}{2}\sigma_{;\mu}\sigma^{;\mu}-U(\sigma)-\frac{1}{4}F^{\mu\nu}F_{\mu\nu}\
,
\end{equation}
with $D_{\mu}\psi\equiv\psi_{;\mu}-\imath eA_{\mu}\psi$, when the
energy-momentum tensor is:
\begin{eqnarray}\label{6.34}
T^{\mu\nu}=\frac{1}{2}\imath(\bar{\psi}\gamma^{(\mu}D^{n)}\psi-
D^{(\mu}\bar{\psi}\gamma^{\nu)}\psi+\sigma^{;\mu}\sigma^{;\nu}-
\nonumber\\
g^{\mu\nu}[(1/2)\sigma_{;\alpha}\sigma^{;\alpha}+U(\sigma)]-
\frac{1}{4}g_{\mu\nu}F^{\alpha\beta}F_{\alpha\beta}+g^{\alpha\beta}
F_{\nu\alpha}F_{\nu\beta}\ .
\end{eqnarray}
Again, regarding only electric fields, defining
$\theta_{\psi}\equiv\omega_{\psi}+eA_0$, and repeating the
discussion of eqs. \ref{5.8}-\ref{5.11} and \ref{5.15}, we finally
have to solve the system of three coupled equations, the two
Einstein:
\begin{equation}\label{6.35}
\frac{1-A}{r^2}-\frac{1}{r}\frac{dA}{dr}=\frac{1}{4\pi}\theta_{\psi}^4B^2
+\frac{\lambda}{4}+\frac{1}{2e^2}\left(\frac{d\theta_{\psi}}{dr}\right)^2+\Lambda\
,
\end{equation}
\begin{equation}\label{6.36}
\frac{A-1}{r^2}-\frac{1}{r}\frac{A}{B}\frac{dB}{dr}=\frac{1}{12\pi}\theta_{\psi}^4B^2
-\frac{\lambda}{4}-\frac{1}{2e^2}\left(\frac{d\theta_{\psi}}{dr}\right)^2-\Lambda\
,
\end{equation}
and the equation of motion for the gauge field:
\begin{eqnarray}\label{6.37}
\frac{d^2\theta_{\psi}}{dr^2}+\left[\frac{2}{r}+\frac{1}{2}
\left(\frac{1}{A}\frac{dA}{dr}+\frac{1}{B}\frac{dB}{dr}\right)\right]\frac{d\theta_{\psi}}{dr}-
\frac{2e^2\langle\bar{\psi}\psi\rangle\theta_{\psi}}{A}&=&0
\Leftrightarrow \nonumber\\
\frac{d^2\theta_{\psi}}{dr^2}+\left[\frac{2}{r}+\frac{1}{2}
\left(\frac{1}{A}\frac{dA}{dr}+\frac{1}{B}\frac{dB}{dr}\right)\right]\frac{d\theta_{\psi}}{dr}-
\frac{2e^2\theta_{\psi}^4B^{3/2}}{A}&=&0\ .
\end{eqnarray}
The total energy of the configuration can be provided by the
$(0,0)$ component of the energy-momentum tensor, when the
particle number is:
\begin{equation}\label{6.38}
Q_{\psi}=4\pi\int
drr^2\sqrt{\frac{1}{A}}\langle\psi^{\dagger}\psi\rangle=
\frac{4}{3\pi}\int drr^2\theta_{\psi}^3B^{3/2}A^{-1/2}\ .
\end{equation}

\section{Concluding remarks}

We studied solitonic field configurations coupled to gravity. We
provided approximate analytical solutions for the scalar field and
solved numerically the Einstein equations for q-stars with one
and two scalar fields, non-abelian q-stars and q-stars with a
fermionic field coupled to a real scalar. The independent
parameters are the cosmological constant, a global spacetime
feature, and the eigen-frequency, with which the soliton rotates
in its internal symmetry space. The main result from the
numerical analysis is that the field at the origin increases
rapidly from the increase, in absolute values, of the cosmological
constant. The mass, radius and particle number, firstly increase
and after a certain value of the cosmological constant decrease.

In the case of charged q-stars, there are stable stars for small
values of the field strength $e$. The soliton parameters take
larger values in the presence of electric charge due to the
electrostatic repulsion between the different parts of the star.
This repulsion has the effect to decrease the $|\phi(0)|$ value
and to force the field concentrate near the surface. For large
values of $e$, the decay into free particles is inevitable.

\vspace{1em}

\textbf{ACKNOWLEDGEMENTS}

\vspace{1em} I wish to thank N.D. Tracas, E. Papantonopoulos and
P. Manousselis for helpful discussions.

\end{document}